\documentclass[aps,showpacs,twocolumn,prl]{revtex4-1}
\usepackage{amssymb}
\usepackage{amsmath}
\usepackage{graphicx}
\usepackage{epsfig}
\usepackage{color}

\begin{document}

\title{Knot Topology in Quantum Spin System}
\author{X. M. Yang}
\author{L. Jin}
\email{jinliang@nankai.edu.cn}
\author{Z. Song}
\email{songtc@nankai.edu.cn}
\affiliation{School of Physics, Nankai University, Tianjin 300071, China}

\begin{abstract}
Knot theory provides a powerful tool for the understanding of topological
matters in biology, chemistry, and physics. Here knot theory is introduced
to describe topological phases in the quantum spin system. Exactly solvable
models with long-range interactions are investigated, and Majorana modes of
the quantum spin system are mapped into different knots and links. The
topological properties of ground states of the spin system are visualized
and characterized using crossing and linking numbers, which capture the
geometric topologies of knots and links. The interactivity of energy bands
is highlighted. In gapped phases, eigenstate curves are tangled and braided
around each other forming links. In gapless phases, the tangled eigenstate
curves may form knots. Our findings provide an alternative understanding of
the phases in the quantum spin system, and provide insights into
one-dimension topological phases of matter.
\end{abstract}

\maketitle

\textit{Introduction.}---Knots are categorized in terms of geometric
topology, and describe the topological properties of one-dimensional (1D)
closed curves in a three-dimensional space \cite{KnotTheory,AB}. A
collection of knots without an intersection forms a link. The significance
of knots in science is elusive; however, knot theory can be used to
characterize the topologies of the DNA structure \cite{DNA} and synthesized
molecular structure \cite{Forgan} in biology and chemistry. Knot theory is
used to solve fundamental questions in physics ranging from microscopic to
cosmic textures and from classical mechanics to quantum physics \cite%
{Wilczek,Rovelli,CNYang,FYWu,Atiyah,VK,Faddeev,Kauffman}. Knots and links
are observed for vortices in fluids \cite{Irvine13,Irvine16}, lights \cite%
{Leach,Dennis}, and quantum eigenfunctions \cite{Taylor}, for solitons in
the Bose--Einstein condensate \cite{Hall}, and for Fermi surfaces of
topological semimetals in Hermitian \cite%
{SCZhang,JMHou,ZWang,ZYan,Ezawa,PYChang,Ackerman,DLD1,DLD2,LLu18,JHu19} and
non-Hermitian systems \cite{JHuPRB,Bergholtz1,Bergholtz2,CHLee}. Knots and
links with distinctive geometric topological features are crucial for
understanding hidden physics.

Innovations in visualization have driven scientific developments.
Historically, Feynman diagrams are created to provide a convenient approach
for describing and calculating complex physical processes in quantum theory
\cite{Feynman}. Feynman diagrams help to understanding the interaction
between particles and provide a thorough understanding of the foundations of
quantum physics. In condensed matter physics, a coherent description of
matter is difficult because of the complexity of its phases. In 1D
topological systems, the phases of matter are generally characterized using
a winding number associated with a Zak phase of the corresponding band.
Topological phases with different winding numbers are considerably different
from each other, and differences in their topological properties can be
attributed to the eigenstates. The winding numbers and Zak phases are
defined for separate energy bands. In the gapless phase, the boundaries of
different gapped phases should have different topological features; however,
the energy band is inseparable. The compatible characterization of gapped
and gapless phases is a concern.

In this Letter, a graphic approach was developed to examine topological
properties and phase transitions in 1D topological systems. The ground
states of the quantum spin system are mapped into knots and links; the
topological invariants constructed from eigenstates are then directly
visualized; and topological features are revealed from the geometric
topologies of knots and links. In contrast to the conventional description
of band topologies, such as the Zak phase and Chern number that can be
extracted from a single band, this approach highlights the interactivity
between upper and lower bands; thus, information on two bands is necessary
for obtaining a complete eigenstate graph. Several standard processes have
been performed to diagonalize the Hamiltonian of the quantum spin system.
The spin-$1/2$ is transformed into a spinless fermion under the
Jordan-Wigner transformation. In Majorana representation, the core matrix of
the spinless fermion system is obtained under Nambu representation. The
eigenstates of the core matrix are then mapped into closed curves of knots
and links. The graphs of different categories represent different
topological phases.

To exhibit various topological phases, an exactly solvable generalized $XY$
model with long-range interactions is revisited, which ensures the richness
of topological phases \cite{GZhang}. The closed curves of eigenstates
completely encode the information on ground states. In topologically
nontrivial phases, eigenstate curves are tangled to form links for the
gapped phase; however, they may be untied and combined into knots for the
gapless phase. Because the winding number (Zak phase) is not appropriately
defined in the gapless phase, knot theory \cite{KnotTheory} is employed to
characterize the topological features of both gapped and gapless phases. In
this approach, the gapped and gapless phases are visualized, and their
topological features are revealed through knots and links.
\begin{figure*}[tb]
\includegraphics[ bb=0 0 490 225, width=17.8 cm,clip]{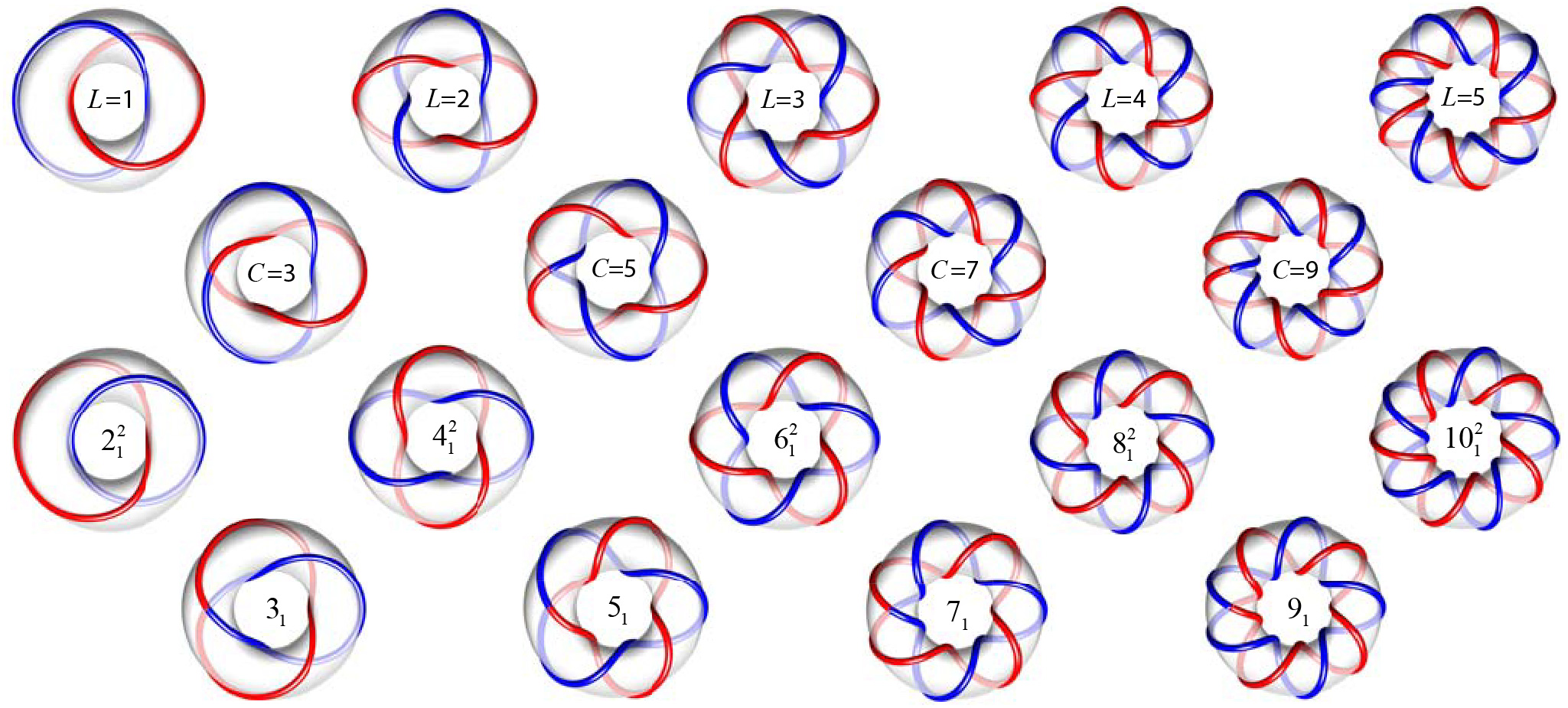}
\caption{The upper (lower) two rows are eigenstate knots and links on the $k$%
-$\protect\varphi $ torus with positive (negative) crossing and linking
numbers. The first and third rows are the links, the second and fourth rows
are the knots. The numbers in the lower two rows indicate the crossing
number, and the superscript represents the number of loops, which is absent
for the single loop knots. The subscript indicates the order of knot
configuration with an identical number of loops and an identical crossing
number in the Alexander-Briggs notation \protect\cite{AB}. The parameters in
Eq.~(\protect\ref{JxJy}) for the $18$ knots and links from left to right and
from top to bottom are $(x,y)$=$(1.1,1.2)$, $(0.5,1.2)$, $(0,1.2)$, $%
(-0.5,1.2)$, $(-0.8,1.2)$; $(0.68,1.2)$, $(0.28,1.2)$, $(-0.28,1.2)$, $%
(-0.68,1.2)$; $(2,0.7)$, $(1.8,0)$, $(1.8,-0.5)$, $(1.8,-1)$, $(1.8,-1.5)$; $%
(2,0.21)$, $(1.5,-0.32)$, $(2,-0.75)$, $(2,-1.2)$.}
\label{fig1}
\end{figure*}

\textit{Quantum spin model.---}A solvable generalized 1D $XY$ model with
long-range three-spin interactions is considered for elucidation \cite%
{Suzuki}. The Hamiltonian is
\begin{equation}
H=\sum\limits_{j=1}^{N}[\sum\limits_{n=1}^{M}\left( J_{n}^{x}\sigma
_{j}^{x}\sigma _{j+n}^{x}+J_{n}^{y}\sigma _{j}^{y}\sigma _{j+n}^{y}\right)
\prod_{l=j+1}^{j+n-1}\sigma _{l}^{z}+g\sigma _{j}^{z}],  \label{H}
\end{equation}%
where operators $\sigma _{j}^{x,y,z}$ are Pauli matrices for spin at the $j$%
th site. $H$ is reduced to an ordinary anisotropic $XY$ model for $M=1$. A
conventional Jordan-Wigner transformation for spin-$1/2$ is performed \cite%
{Sachdev}, and the Hamiltonian of a subspace with odd number particles can
be expressed using a Hamiltonian describing a spinless fermion as follows
\begin{equation}
H=\sum\limits_{j=1}^{N}[\sum\limits_{n=1}^{M}(J_{n}^{+}c_{j}^{\dag
}c_{j+n}+J_{n}^{-}c_{j}^{\dag }c_{j+n}^{\dag }+\mathrm{H.c.}%
)+g(1-2c_{j}^{\dagger }c_{j})],  \label{H_f}
\end{equation}%
where $c_{j}^{\dagger }$ ($c_{j}$) denotes the creation (annihilation)
operator of spinless fermion at the $j$th site and $J_{n}^{\pm
}=J_{n}^{x}\pm J_{n}^{y}$. In large $N$ limit ($N\gg M$), $H$\ is available
for both parities of particle numbers, thus representing a $p$-wave
topological superconducting wire using long-range interactions \cite%
{kitaev,Niu}. Majorana fermion operators are introduced, $%
a_{j}=c_{j}^{\dagger }+c_{j},b_{j}=-i(c_{j}^{\dagger }-c_{j})$, the inverse
transformation gives $c_{j}^{\dagger }=\left( a_{j}+ib_{j}\right)
/2,c_{j}=\left( a_{j}-ib_{j}\right) /2$. The Majorana representation of the
Hamiltonian is%
\begin{equation}
H=i\sum\limits_{j=1}^{N}[\sum\limits_{n=1}^{M}\left(
J_{n}^{x}b_{j}a_{j+n}-J_{n}^{y}a_{j}b_{j+n}\right) +ga_{j}b_{j}].
\end{equation}%
In the Nambu representation of basis $\psi ^{\mathrm{T}}=(a_{1}$, $ib_{1}$, $%
a_{2}$, $ib_{2}$, $...)$, $H=\psi ^{\mathrm{T}}h\psi $ and $h$\ is a $%
2N\times 2N$ matrix. Applying Fourier transformation, we obtain $%
h=\sum_{k}h_{k}$ and
\begin{equation}
h_{k}=B_{x}\sigma _{x}+B_{y}\sigma _{y}\mathbf{,}
\end{equation}%
where $\sigma _{x}=\left(
\begin{array}{cc}
0 & 1 \\
1 & 0%
\end{array}%
\right) $ and $\sigma _{y}=\left(
\begin{array}{cc}
0 & -i \\
i & 0%
\end{array}%
\right) $. The components of the effective magnetic field are periodic
functions of momentum $k$
\begin{equation}
B_{x}=\sum\limits_{n=1}^{M}J_{n}^{+}\cos \left( nk\right)
/4-g/4,B_{y}=\sum\limits_{n=1}^{M}J_{n}^{-}\sin \left( nk\right) /4,
\label{B}
\end{equation}%
all information on the system is encoded in matrix $h_{k}$, because the
eigenvalues and eigenvectors can be employed to construct the complete
eigenstates of the original Hamiltonian. A winding number is defined using
an effective planar magnetic field $\left( B_{x},B_{y}\right) $ as follows
\begin{equation}
w=\frac{1}{2\pi }\oint_{C}\mathbf{(}\hat{B}_{x}\nabla \hat{B}_{y}-\hat{B}%
_{y}\nabla \hat{B}_{x}),  \label{w}
\end{equation}%
where $\hat{B}_{x(y)}=B_{x(y)}/B$, $B=\left( B_{x}^{2}+B_{y}^{2}\right)
^{1/2}$, and $C$ is the Bloch vector curves in the $B_{x}$-$B_{y}$ plane. $w$
characterizes a vortex enclosed in the loop. In the gapless phase, loop $C$
passes through the origin, and integral $w$ depends on the loop path.
Although $w$ is crucial for characterizing the gapped phase, $w$ cannot be
used to characterize the gapless phase (see Supplementary Material A~\cite%
{SI}).

To overcome this concern, a graphic representation is developed to visualize
the eigenstates (Fig.~\ref{fig1}). The eigenstates in different gapless and
gapped phases are mapped into different categories of torus knots and links.
The topological features of the ground state are evident through the
geometric topologies of eigenstate knots and links on $k$-$\varphi $ torus.
Therefore, knot theory in mathematics can characterize topology phases in
physics. The obtained knots and links depend on Fourier transformation.
However, different topological phases are still distinguishable in the
graphic representation; moreover, $h_{k}$ and knots are unique for the set
Fourier transformation. In the following sections, we discuss the graphic
eigenstate, topological characterization, and knot theory for the
topological phases of the quantum spin system.

The spectrum of $h_{k}$ is $\epsilon _{k}^{\pm }=\pm B$ and the
corresponding eigenstate is $\left\vert \psi _{k}^{\pm }\right\rangle
=[e^{i\varphi _{\pm }(k)},1]^{\mathrm{T}}/\sqrt{2}$. The phases in
eigenstate $\left\vert \psi _{k}^{\pm }\right\rangle $ are real periodic
functions of $k$, $\varphi _{+}(k)=\arctan \left( -B_{y}/B_{x}\right) $, and
\begin{equation}
\varphi _{-}(k)=\varphi _{+}(k)+\pi .
\end{equation}%
The ground state phase diagram is obtained using $\epsilon _{k_{c}}^{\pm }=0$%
, where the energy band gap closes.

\textit{Knot topology.---}The topological features of the ground state are
completely encoded in phase factor $\varphi _{\pm }(k)$. In the absence of
band degeneracy, the energy bands are gapped and each eigenstate is
represented as a closed curve on the surface of an unknotted torus in $%
\mathbf{R}^{3}$. The toroidal direction is $k$ and the poloidal direction is
$\varphi $ (Fig.~\ref{fig1}). From $\varphi _{-}(k)=\varphi _{+}(k)+\pi $,
the two curves are always located at opposite points in the cross section of
the torus. For topologically nontrivial phases, they become tangled curves,
which are braided around each other. In the presence of band degeneracy, two
eigenstates may form one closed curve. The topological properties of the
ground state are reflected from the geometric topology of closed curves. The
curves on the torus forms a torus knot, which is a particular type of
knots/links on the surface of an unknotted torus \cite{KnotTheory}. For the
eigenstate graphs of the corresponding quantum spin system, eigenstate
graphs are only single-loop knots and two-loop links. The rich topological
phases of either gapless or gapped, either Hermitian or non-Hermitian band
degeneracy, and either trivial or nontrivial phases are all distinguishable
from the eigenstate graphs. The graphic approach is employed to distinguish
between the real gapless phases arising from exceptional points and
degenerate points, which cannot be distinguished using the two winding
numbers in the non-Hermitian system \cite{YXM}.

For a gapped system, the two loops without any node form a link on the
torus. In knot theory, for the topology of the two closed curves (loops) in
the three-dimensional space, a linking number is used as a topological
invariant \cite{Alexander}. The linking number represents the number of
times each curve is braided around the other curve. Mathematically, the
linking number of two closed curves $\mathbf{r}_{+}(k)$\ and $\mathbf{r}%
_{-}(k^{\prime })$\ can be calculated using a double line integral%
\begin{equation}
L=\frac{1}{4\pi }\oint\nolimits_{k}\oint\nolimits_{k^{\prime }}\frac{\mathbf{%
r}_{+}(k)-\mathbf{r}_{-}(k^{\prime })}{\left\vert \mathbf{r}_{+}(k)-\mathbf{r%
}_{-}(k^{\prime })\right\vert ^{3}}\cdot \lbrack \mathrm{d}\mathbf{r}%
_{+}(k)\times \mathrm{d}\mathbf{r}_{-}(k^{\prime })].  \label{L1}
\end{equation}%
The two curves are the two loops of eigenstates; $\left\vert \mathbf{r}%
_{+}\left( k\right) -\mathbf{r}_{-}\left( k^{\prime }\right) \right\vert $
for $k=k^{\prime }$ is the cross section diameter of the torus. A
straightforward derivation yields (see Supplementary Material B~\cite%
{SI,Asboth})%
\begin{equation}
L=-\frac{1}{2\pi }\int_{0}^{2\pi }\nabla _{k}\varphi _{+}\left( k\right)
\mathrm{d}k,  \label{L2}
\end{equation}%
which has a geometrical meaning: the braiding of the two curves on the torus
surface. Furthermore, $L=w$ is valid for gapped phases (see Supplementary
Material B~\cite{SI}); and linking number $L$ for the two loops has the same
significance as that of winding number $w$ in the characterization of the
topology of the ground state.

The gapless phase is the phase transition boundary of different gapped
phases. The band degenerate points in a chiral symmetric system have zero
energy. At degenerate points, eigenstates switch and experience a $\pi $
phase shift in both phase factors $\varphi _{\pm }\left( k\right) $ because $%
\varphi _{+}\left( k\right) $ and $\varphi _{-}\left( k\right) $ have a
phase difference of $\pi $. In the period $[-\pi ,\pi ]$ of $k$, $-\pi $ and
$\pi $ are considered as one point degenerate point. For the two energy
bands with an even number of degenerate points in a $2\pi $ period of $k$.
The energy bands are repeated in the subsequent period of $k$, and the two
corresponding separate eigenstate loops are observed. By contrast, for two
energy bands with an odd number of degenerate points in a $2\pi $ period of $%
k$, the two energy bands are switched in the subsequent $2\pi $ period of $k$
as they switched odd times because of band degeneracies. In this case, two
functions $\varphi _{+}(k)$ and $\varphi _{-}(k)$ combine to form a single
periodic function and the corresponding eigenstate curves are connected to
form a single-loop on the torus. In contrast to a two-loop link, the graph
eigenstates comprise a knot.

Alternatively, considering the $\pi $ phase shift in $\varphi _{\pm }\left(
k\right) $ at the band degenerate points, a total phase change is $2\pi $
for two eigenstates. When $k$ varies by $2\pi $, in any topological phase
with a two-loop link, phase $\varphi _{\pm }\left( k\right) $ is an integer
of $2\pi $ and the total circling $\Phi \left( k\right) =\varphi _{+}\left(
k\right) +\varphi _{-}\left( k\right) $ is an even times of $2\pi $.
Therefore, the appearance of an odd number of degenerate points in the
energy band results in the change of an odd times of $2\pi $. Thus, total
circling $\Phi \left( k\right) $ becomes an odd times of $2\pi $, the
two-loop link must change into a single-loop knot. A topological phase with
an even number of band degenerate points is represented by a two-loop link.
Thus, the band degenerate induces the transition of eigenstate graph between
the link and knot.

Knot topology is characterized by using crossing number $c\left( K\right) $,
which is defined as a minimal number of loop intersections in any planar
representation \cite{KnotTheory}, and $K$ in the bracket represents the
knot. The transition between a link and knot, and between two different
links/knots with different linking/crossing numbers are only achieved by
untying two closed curves and by alternatively reconnecting the endpoints.
Continuous deformation cannot change the type of knots and links. This
observation indicates the robustness of the ground state provided by
topological protection.

All knots represent gapless phases. Considering the gapless phase as the
boundary of two gapped phases $A$ and $B$, if the linking numbers of two
gapped phases have an even difference, and the in-between gapless phase has
an even number of degenerate points, the two bands are mixed because of the
band degeneracy and switch even times. The gapless phase is represented by a
link with the following linking number $L=(L_{A}+L_{B})/2$. If the linking
numbers of the two gapped phases have an odd difference, and the in-between
gapless phase has an odd number of degenerate points, the gapless phase is
represented by a knot, and the crossing number is $c\left( K\right)
=L_{A}+L_{B}$. In knot representation, a gapless phase with an even number
of band degenerate points never has the same linking number as that of a
gapped phase on either side of the gapless phase. Thus, all types of
topological phases are directly distinguishable based on the knot topology
of eigenstate graphs.

\textit{1D }$\mathit{XY}$\textit{\ model.---}To demonstrate the rich
topological features of ground states in different gapped and gapless
phases, we consider a quantum spin model with the interactions $\left(
J_{n}^{x},J_{n}^{y}\right) $ and the transverse field $g=0$. To observe
topological phases with high winding numbers, a large $M$ is required in
Hamiltonian $H$, and the system has long-range interactions. We consider the
interaction between spins exponentially decay as their distances. As an
example, the interactions are set%
\begin{equation}
\begin{array}{l}
J_{n}^{x}=\exp [-\left( 2x+n-3\right) ^{2}], \\
J_{n}^{y}=2\left( y-1\right) \exp [-\left( 2y+n-2\right) ^{2}],%
\end{array}
\label{JxJy}
\end{equation}%
where $x,y$ determine the strengths of interactions. Rich topological phases
are obtained because of long-range interactions. The boundary of different
phases is obtained by calculating the band gap closing condition, i.e., $%
\epsilon _{k_{c}}^{\pm }=0$. The corresponding winding number can be
obtained through a numerical integration of Eq. (\ref{w}).

Alternatively, the eigenstate graph provides a clear picture and a
convenient approach for identifying the topological properties of different
ground states in the spin system. The topological properties of the ground
states are revealed using the geometric topologies of eigenstate knots and
links. The geometric topologies of knots and links only change for the
untying and reconnecting of the components (loops), which help understand
topological phase transition in the quantum spin system. The eigenstate
knots and links of different topological phases are shown in Fig.~\ref{fig1}.

The gapped phase corresponds to a link, where the linking number is equal to
the corresponding winding number ($L=w$). In the first row of Fig.~\ref{fig1}%
, the topological phases with winding numbers $w=1$ to $5$ are elucidated
through their eigenstate graphs with $L=1$ to $5$, respectively. $L=0$ (not
shown) is unknotted and corresponds to a topologically trivial phase.
Moreover, with $k$ increasing, two loops are rotated clockwise
(counterclockwise) in the cross section of the torus if the linking number
of the graph is positive (negative). The graph in Fig.~\ref{fig1} with $L=1$
shows a Hopf link, with a winding number of $w=1$ represented by $2_{1}^{2}$%
. The graph with $L=2$ presents a Solomon's knot, it belongs to a link in
contrast to its name and its notation is $4_{1}^{2}$. The graph with $L=3$
is the star of David denoted as $6_{1}^{2} $.

The gapless phase separates two gapped phases with linking numbers $L_{A}$
and $L_{B}$, and its eigenstate graph corresponds to a link with a linking
number of $L=\left( L_{A}+L_{B}\right) /2$ for two degenerate points in the
energy band. The graphs are similar to those in the first and third rows of
Fig.~\ref{fig1}.

The gapless phase separating the gapped phases with linking numbers $L_{A}$\
and $L_{B}$ corresponds to a knot with a crossing number of $c\left(
K\right) =L_{A}+L_{B}$ for one band degenerate point. With the increasing $k$%
, the point on the knot is rotated clockwise (counterclockwise) in the cross
section of the torus if the crossing number of the graph is positive
(negative). The second and fourth rows in Fig.~\ref{fig1} show knots that
separate gapped phases in the first and third rows, where each knot denotes
the gapless phase between the two gapped phases represented by the two
nearest aforementioned links. The magnetic nonsymmorphic symmetry of $h$
ensures the band degeneracy at $\pi $ \cite{YXZhao}. The gapless phase with
crossing number of $c\left( K\right) =3$ has a positive trefoil knot, and it
separates gapped phases with $L_{A}=1$\ and $L_{B}=2$. The corresponding
trefoil knot in the fourth row marked with $3_{1}$ is a negative trefoil
knot, and its crossing number is $c\left( K\right) =-3$. A
pentafoil/cinquefoil knot with a crossing number of $c\left( K\right) =5$
separates the gapped phases with $L_{A}=2$\ and $L_{B}=3$, the $c\left(
K\right) =-5$ knot is marked with $5_{1}$. The knot with crossing number $%
c\left( K\right) =7$ [$c\left( K\right) =9$] represents the gapless phase,
which separates the gapped phases with $L_{A}=3$ and $L_{B}=4$ ($L_{A}=4$
and $L_{B}=5$). The number of knot windings on the torus is $c\left(
K\right) $ and has $c\left( K\right) $ number of crossing points in a planar
plane.

\textit{Conclusion.---}We introduce a graphic approach to investigate the
topological phase transition in the quantum spin system. The eigenstates
completely encode topological features, which are vividly revealed from the
eigenstate curves, tangled and braided around each other into knots and
links. The geometric topologies of knots and links represent the topological
properties of different ground states, and characterize the topologies of
both the gapped and gapless phases. The graphic approach highlights the
interactivity of the energy bands. Our findings provide insights into the
application of knot theory in the quantum spin system and topological phase
of matter.

\textit{Acknowledgement.---}This work was supported by National Natural
Science Foundation of China (Grants No.~11874225 and No.~11605094).

\clearpage

\newpage
\begin{widetext}

\section*{Supplemental Material for "Knot topology in quantum spin system"}

\begin{center}
X. M. Yang, L. Jin, and Z. Song\\[2pt]
\textit{School of Physics, Nankai University, Tianjin 300071, China}

\bigskip

\textbf{A: Bloch vector and winding number}
\end{center}

The planar Bloch vector in the plane $B_{x}$-$B_{y}$ for the situations
shown in Fig.~1 of the main text is depicted in Supplementary Figure~\ref%
{figS1}. The winding of Bloch vector around the origin (black dot) of $B_{x}$%
-$B_{y} $ plane in the gapped phase is $w$. In the first and third rows, the
curve of Bloch vector encloses the origin one to five times from the left to
right, and the winding number in Eq.~(6) of the main text is $w=\pm 1$ to $w=\pm 5$, respectively. In
the second and fourth rows, the curve of Bloch vector passes through the
origin. The arrows indicate the direction of the curves as $k$ increasing
from $0$ to $2\pi$; the counterclockwise (clockwise) direction yields the $+$
($-$) sign of the winding number.

\begin{figure}[h]
\includegraphics[ bb=0 0 560 350, width=18.0 cm, clip]{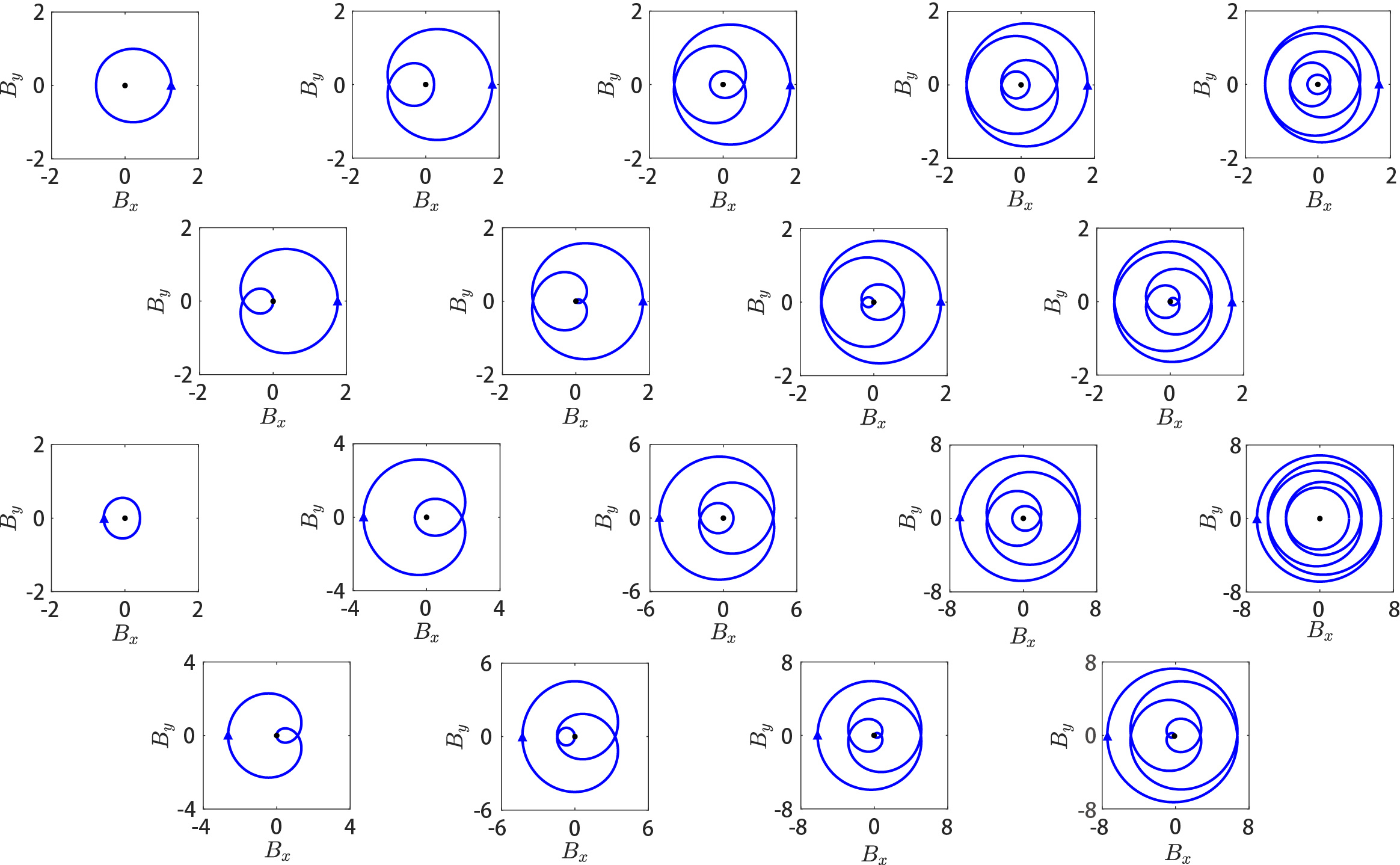}
\caption{Curve of Bloch vector in the $B_{x}$-$B_{y}$ plane, the black
dot is the origin. The corresponding knot or link eigenstate graph is depicted in Fig. 1
of the main text.} \label{figS1}
\end{figure}

\begin{center}
\textbf{B: Linking number}
\end{center}

In the gapped phase, the eigenstates $\left\vert \psi _{k}^{\pm
}\right\rangle $ are represented by two loops on a torus surface, forming a
torus link [Supplemental Figure~\ref{figS2}(a)]. $\mathbf{r}_{+}\left(
k\right) $ (blue loop) and $\mathbf{r}_{-}\left( k^{\prime }\right) $ (red
loop) represent the positive branch $\left\vert \psi _{k}^{+}\right\rangle $
and the negative branch $\left\vert \psi _{k^{\prime }}^{-}\right\rangle $,
respectively. We construct a vector field%
\begin{equation}
\mathbf{h}\left( k,k^{\prime }\right) =\mathbf{r}_{+}\left( k\right) -%
\mathbf{r}_{-}\left( k^{\prime }\right) ,
\end{equation}%
where $\mathbf{r}_{+}\left( k\right) $ and $\mathbf{r}_{-}\left( k^{\prime
}\right) $ in the cylindrical coordinate [Supplemental Figure~\ref{figS2}%
(b)] are expressed as
\begin{eqnarray}
\mathbf{r}_{+}\left( k\right) &=&0\bm{\hat{\theta} }+r\sin \varphi
_{+}\left( k\right) \mathbf{\hat{z}}+\left[ R+r\cos \varphi _{+}\left(
k\right) \right] \mathbf{\hat{r}},  \notag \\
\mathbf{r}_{-}\left( k^{\prime }\right) &=&0\bm{\hat{\theta} }^{\prime
}+r\sin \varphi _{-}\left( k^{\prime }\right) \mathbf{\hat{z}}^{\prime }+%
\left[ R+r\cos \varphi _{-}\left( k^{\prime }\right) \right] \mathbf{\hat{r}}%
^{\prime }.
\end{eqnarray}%
On the torus, $R$ (in red) is the distance from the center of the tube to
the center of the torus, and $r$ (in purple) is the radius of the tube. Two
cylindrical coordinates are related as follows%
\begin{equation}
\mathbf{\hat{z}}^{\prime }=\mathbf{\hat{z},\hat{r}}^{\prime }=\cos \left(
k^{\prime }-k\right) \mathbf{\hat{r}}+\sin \left( k^{\prime }-k\right) %
\bm{\hat{\theta} }.
\end{equation}%
Besides, we have $\varphi _{-}\left( k\right) =\varphi _{+}\left( k\right)
+\pi $ and $\varphi _{-}\left( k^{\prime }\right) =\varphi _{+}\left(
k^{\prime }\right) +\pi .$ Based on these relations, we have
\begin{equation}
\mathbf{h}\left( k,k^{\prime }\right) =h_{\theta }\bm{\hat{\theta} }+h_{z}%
\mathbf{\hat{z}}+h_{r}\mathbf{\hat{r}},
\end{equation}%
where
\begin{eqnarray}
h_{\theta } &=&\left[ R-r\cos \varphi _{+}\left( k^{\prime }\right) \right]
\sin \left( k-k^{\prime }\right) ,  \notag \\
h_{z} &=&r\sin \varphi _{+}\left( k\right) +r\sin \varphi _{+}\left(
k^{\prime }\right) , \\
h_{r} &=&R+r\cos \varphi _{+}\left( k\right) -\left[ R-r\cos \varphi
_{+}\left( k^{\prime }\right) \right] \cos \left( k-k^{\prime }\right) .
\notag
\end{eqnarray}%
Each point $\left( k,k^{\prime }\right) $ is mapped to a three-dimensional
vector $\mathbf{h}$. Since $k$ and $k^{\prime }$ are periodic parameters,
the endpoint of vector $\mathbf{h}$ draws a deformed torus surface in the $%
\left( \bm{\theta },\mathbf{z},\mathbf{r}\right) $ space.

\begin{figure}[h]
\includegraphics[bb=100 30 690 205, width=14 cm, clip]{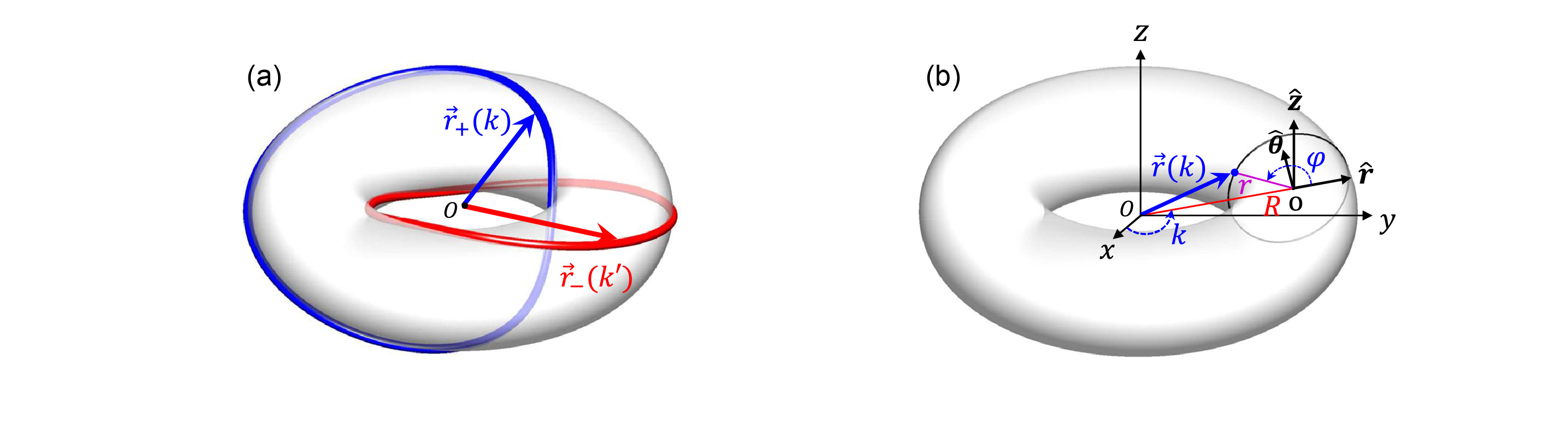}
\caption{(a) A two-component Hopf link lies on the torus surface, each component is a
closed loop of the eigenstate $\left\vert \protect\psi _{k}^{\pm
}\right\rangle $. (b) The cylindrical coordinate for the torus.} \label%
{figS2}
\end{figure}

After introducing the vector field $\mathbf{h}$, we note that the linking
number of two curves $\mathbf{r}_{+}\left( k\right) $ and $\mathbf{r}%
_{-}\left( k^{\prime }\right) $ in Eq. (8) of the main text can be expressed
in the form of%
\begin{equation}
L=\frac{1}{4\pi }\oint\nolimits_{k}\oint\nolimits_{k^{\prime }}\frac{\mathbf{%
h}}{\left\vert \mathbf{h}\right\vert ^{3}}\cdot (\frac{\partial \mathbf{h}}{%
\partial k}\mathrm{d}k\times \frac{\partial \mathbf{h}}{\partial k^{\prime }}%
\mathrm{d}k^{\prime }).
\end{equation}%
Mathematically, the above expression of linking number is equal to the solid
angle extended by the deformed torus surface dividing by $4\pi $ \cite%
{Asboth} and the solid angle can be considered as twice the plane angle of
the curve in the cross section of the deformed torus surface, whereas the
cross section must contain the origin $\left( 0,0,0\right) $ in the $\left( %
\bm{\theta },\mathbf{z},\mathbf{r}\right) $ space.

To get the cross section containing the origin, we set $h_{\theta }=0$. The $%
h_{\theta }=0$ plane cuts off the deformed torus surface to a curve in the $%
\left( \mathbf{z,r}\right) $ plane. By setting $h_{\theta }=0,$ we have $%
k^{\prime }=k$ and $\mathbf{h}$ is reduced to
\begin{equation}
\mathbf{h}=h_{z}\mathbf{\hat{z}}+h_{r}\mathbf{\hat{r}}=2r\sin \varphi
_{+}\left( k\right) \mathbf{\hat{z}+}2r\cos \varphi _{+}\left( k\right)
\mathbf{\hat{r}},
\end{equation}%
which indicates a closed circle $\Gamma $ centered at $\left( 0,0\right) $
in the $\left( \mathbf{z,r}\right) $ plane and the linking number defined by
the vector field $\mathbf{h}$ is given by%
\begin{equation}
L=\frac{1}{2\pi }\oint_{\Gamma }\frac{1}{4r^{2}}\mathbf{(}h_{z}\nabla
h_{r}-h_{r}\nabla h_{z})=-\frac{1}{2\pi }\int_{0}^{2\pi }\nabla _{k}\varphi
_{+}\left( k\right) \mathrm{d}k.  \label{L}
\end{equation}%
Furthermore, from the definition $\varphi _{+}\left( k\right) =\arctan
\left( -B_{y}/B_{x}\right) $, the linking number Eq. (\ref{L}) can be
expressed as%
\begin{equation}
L=\frac{1}{2\pi }\oint_{C}\frac{1}{B^{2}}\mathbf{(}B_{x}\nabla
B_{y}-B_{y}\nabla B_{x})=w.
\end{equation}%
The linking number $L$ is equivalent to the winding number $w$ defined in
Eq.~(6) of the main text.
\clearpage
\end{widetext}

\end{document}